\documentclass[a4paper,11pt]{article}

\usepackage{amssymb}
\usepackage{fullpage}

\def\01{\{0,1\}}

\def\EQP{\mathsf{EQP}}
\def\BQP{\mathsf{BQP}}
\def\BPP{\mathsf{BPP}}
\def\MOD2P{\mathsf{MOD_2P}}

\def\MODmP{\mathsf{MOD_mP}}
\def\MODpkP{\mathsf{MOD_{p^k}P}}
\def\PARP{\mathsf{\oplus P}}
\def\PNP{\mathsf{P^{NP}}}
\def\NPintCONP{\mathsf{NP \cup coNP}}
\def\PH{\mathsf{PH}}
\def\S2P2{\mathsf{\Sigma_2^p \cap \Pi_2^p}}
\def\SI2{\mathsf{\Sigma_2^p}}

\def\etal{\textit{et al.~}}

\def\GF2{\mathbb{GF}(2)}

\def\mon{\mathrm{mon}}

\newcommand{\url}[1]{\texttt{#1}}

\newcommand{\pr}[1]{\mathrm{Pr}[ #1 ]}

\newtheorem{Theorem}{Theorem}
\newtheorem{Lemma}[Theorem]{Lemma}
\newtheorem{Definition}{Definition}
\newtheorem{Corollary}[Theorem]{Corollary}

\newenvironment{Proof}{%
  \begin{trivlist}{}{\setlength{\topsep}{0cm}\setlength{\partopsep}{0cm}}
  \item {\bf Proof\@}\hspace*{1ex}\ignorespaces}%
  {\makebox[0cm]{}\nolinebreak\hfill$\Box$\end{trivlist}}

\title{Comparing $\EQP$ and $\MODpkP$ using\\ Polynomial Degree Lower Bounds}
\author{Mart~de~Graaf\thanks{CWI.
P.O.~Box 94079, 1090 GB Amsterdam, The Netherlands. Email:
$\mathtt{mgdgraaf@cwi.nl}$. Partially supported by the EU fifth
framework project QAIP, IST--1999--11234, and by grant 612.055.001
from the Netherlands Organization for Scientific Research (NWO).}
\and Paul~Valiant\thanks{Stanford University P.O. Box 17308, 
Stanford CA 94309.  Email: $\mathtt{pvaliant@stanford.edu}$.
Research supported in part by the DoD Multidisciplinary University
Research Initiative (MURI) program administered by the Army Research
Office under grant DAAD19-00-1-0177.}}

\begin{document}
\maketitle

\begin{abstract}
We show that an oracle $A$ that contains either $1/4$ or $3/4$ of all strings of length $n$ can be used to separate $\EQP$ from 
the counting classes $\MODpkP$, where $p$ is a prime. Our proof makes use of the degree of 
a representing polynomial over $\mathbb{Z}_{p^k}$. We show a linear lower bound on the degree 
of this polynomial. We also show an upper bound of $\mathcal{O}(n^{1/\log_p m})$ on the degree over 
the ring $\mathbb{Z}_m$, whenever $m$ is a squarefree composite with largest prime factor $p$.\end{abstract}

\section{Introduction}
One of the central goals of complexity theory is to understand the various relationships
between complexity classes. In particular, with the introduction of \emph{quantum} complexity
theory, an exciting new challenge has arisen in understanding the relationship
between classical and quantum classes. In particular, one asks about the strength
of $\BQP$, the class of all problems that can be efficiently solved using a quantum computer with
bounded error, and $\EQP$, the class of problems that can be efficiently solved using a quantum
computer which always gives the right answer, compared to classical complexity classes. Unfortunately,
questions in this direction are notoriously hard to settle.

A more feasible task however, is to show that relative to some oracle, a certain relationship
between two complexity classes holds. Early results include a relativized
separation of $\BQP$ from $\BPP$ by Bernstein and Vazirani \cite{bernstein&vazirani:qcomplexity},
and a relativized separation of $\EQP$ from $\NPintCONP$ by Berthiaume and Brassard \cite{bb:oracle}.
Green and Pruim \cite{green&pruim:eqp_pnp} improved upon the latter result by exhibiting 
an oracle relative to which $\EQP \nsubseteq \PNP$.

In this paper we ask whether $\EQP$ can be separated from $\MODmP$ by an oracle. Note that due the linear lower
bound on the degree of a polynomial representing the parity function over the reals  
(Beals \etal \cite{bbcmw:polynomials}), the other direction, separating $\MODmP$ from $\EQP$, is easy.
Recall that $\MODmP$
is the class of languages decided by non-deterministic polynomial time machines that accept iff the number 
of accepting computation paths is nonzero modulo $m$. 
In particular, we ask whether an oracle that is promised to hold either $1/4$ or $3/4$ of all the strings
of each length can be used to separate $\EQP$ from $\MODmP$. This leads us to 
investigate the degree of a polynomial $q:\mathbb{Z}_m^n\to\mathbb{Z}_m$ that for $x\in\01^n$ has 
$q(x)\neq0$ if $|x|=n/4$, and $q(x)=0$ if $|x|=3n/4$, where $|x|$ denotes the number of 1's in a binary string.
When $m$ is a prime power, we show a linear lower bound on the degree of any such polynomial.
This implies that for prime $p$, $\EQP$ can be separated from $\MODpkP$ (and specifically $\PARP$).
We then focus our attention on composite $m$. If $m$ is a squarefree composite, we show an upper bound 
on the degree of $\mathcal{O}(n^{1/\log_pm})$, where $p$ is the largest prime factor of $m$. As is the case with 
the $OR$ function (Barrington, Beigel, and Rudich \cite{bbr:modulo}), this gives another example of a Boolean 
function whose representing degree drops significantly if we go from prime power moduli to composite moduli.

\section{Preliminaries}
\subsection{Complexity Theory}
We assume familiarity with the basics of classical and quantum complexity
theory. For the former \cite{papadimitriou:cc} provides an excellent introduction,
for the latter we recommend \cite{nielsen&chuang:qc}. In particular we are
interested in the complexity classes $\MODmP$ and $\EQP$, definitions of which
are provided here for completeness' sake. Let $M$ be a non-deterministic Turing machine.
By $\#M(x)$ we denote the number of accepting computations of $M$ on input $x$.
\begin{Definition}Let $L\subseteq\01^*$. We say that $L \in \MODmP$ iff there exists 
a polynomial time non-deterministic Turing machine $M$, such that
\begin{enumerate}
\item $x\in L \Rightarrow \#M(x)\bmod m\neq 0$
\item $x\notin L \Rightarrow \#M(x)\bmod m= 0$
\end{enumerate}
\end{Definition}
\begin{Definition}Let $L\subseteq\01^*$. We say that $L \in \EQP$ iff there exists 
a polynomial time quantum Turing machine $M$, such that
\begin{enumerate}
\item $x\in L \Rightarrow \pr{\textrm{M accepts x}}=1$
\item $x\notin L \Rightarrow \pr{\textrm{M accepts x}}=0$
\end{enumerate}
\end{Definition}
We define relativized versions of these complexity classes in the usual way.
\subsection{Combinatorics}
For natural numbers $n$ and $k$, we denote by $(n)_k$ the $k$-ary representation
of $n$, i.e. the string $\ldots a_2a_1a_0$, with $0 \leq a_i < k$, such that 
$n=\sum_i a_ik^i$. Note that the first (from the right) nonzero digit of $(n)_k$ is 
given by the least $i$ such that $k^i \nmid n$, an observation to which we shall frequently refer.

In 1878 Lucas \cite{lucas:congruences} gave a method to easily determine the value 
of ${n \choose k} \bmod p$, for prime $p$, and the following theorem is now known as 
Lucas' Theorem. It is one of the main ingredients in the proofs of our results. 
By $x[i]$ we denote the symbol at the $i$th position of string $x$.
\begin{Theorem}[Lucas]\label{thm:lucas}Let $p$ be a prime number, and $n,k$ positive
integers, then
$${n \choose k} \bmod p = \prod_{i=1}^m {(n)_p[i] \choose (k)_p[i]} \bmod p,$$
where $m$ is the maximal index $i$ such that $(n)_p[i]\neq0$ or $(k)_p[i]\neq0$,
and we use the convention that ${0 \choose x}=0$ whenever $x>0$.
\end{Theorem}
Another theorem that we shall make use of in this paper is the Chinese Remainder
Theorem. We state it here for completeness' sake.
\begin{Theorem}[Chinese Remainder Theorem]\label{thm:crt}Let $r_1,\ldots,r_\ell$ be
pairwise relatively prime, and $m=\prod_{i=1}^\ell r_i$. Then 
$$\mathbb{Z}_m \cong \mathbb{Z}_{r_1} \times \ldots \times \mathbb{Z}_{r_\ell},$$
where the isomorphism is given by $\psi(x \bmod m) \mapsto (x \bmod r_1,\ldots,x \bmod r_\ell)$.\end{Theorem}

\subsection{Representation of Boolean Functions over $\mathbb{Z}_m$}
We now define what it means for a polynomial over $\mathbb{Z}_m$ to represent a Boolean
function. We should note that there are different opinions on what would be the most natural
definition of representing a Boolean function by a polynomial over $\mathbb{Z}_m$,
see for instance the discussion in Tardos and Barrington \cite{tardos:lower}. The definition
we use here, is what is sometimes called \emph{one-sided representation}.

\begin{Definition}Let $g:\01^n\to\01$ be a Boolean function, and $p:\mathbb{Z}_m^n\to
\mathbb{Z}_m$ a polynomial. We say that \it{$p$ represents $g$ over $\mathbb{Z}_m$} 
iff for all $x \in\01^n$, $p(x)=0 \Leftrightarrow g(x)=0$.
By the degree $\deg(p)$ of a polynomial $p:\mathbb{Z}_m^n\to\mathbb{Z}_m$, we mean the 
size of its largest monomial. The degree of a Boolean function $g:\01^n\to\01$ over
$\mathbb{Z}_m$ is then defined as $\deg(g,m) = \min\{\deg(p) \mid \textrm{$p$ 
represents $g$ over $\mathbb{Z}_m$}\}$.\end{Definition}

Note that since for all $x\in\01$ and $\ell>0$, we have that $x^\ell=x$, 
we can restrict ourselves to \emph{multilinear} polynomials.

When the modulus is a prime, we have the following two interesting lemmas. Both are usually
stated as being folklore results. See \cite{beigel:poly} for an overview of these and 
other similar results. 
\begin{Lemma}\label{lem:pol01}Let $p$ be a prime, and $g:\mathbb{Z}_p^n\to\mathbb{Z}_p$ be 
a polynomial of degree $d$, then there is a polynomial $h:\mathbb{Z}_p^n\to\mathbb{Z}_p$ 
of degree $(p-1)d$, such that for all $x \in \01^n $, $h(x)\in\01$, and $h(x) = 0$ 
iff $g(x)=0$.\end{Lemma}
\begin{Proof}Take $h=g^{p-1}$. By Fermat's little theorem, 
$h(x)\equiv1\bmod p$ iff $g(x)\neq0$.\end{Proof}
We should note that Theorem 19, item (ii) in \cite{beigel:poly} contains an erroneous 
proof. We have learned about a correct result via Richard Beigel (personal communication, 
October 2002). It is stated in the next lemma.
\begin{Lemma}\label{lem:ppk}Let $k$ be a positive integer, and $p$ a prime. If 
$g:\mathbb{Z}_{p^k}^n \to \mathbb{Z}_{p^k}$ is a polynomial of degree $d$, then
there exists a degree $d(2p^{k-1}-1)$ polynomial $h:\mathbb{Z}_p^n\to\mathbb{Z}_p$,
such that for all $x\in\01^n$, $h(x)=0$ iff $g(x) = 0$.\end{Lemma}
\begin{Proof}By Theorem \ref{thm:lucas}, we have that for every prime $p$, and positive
integer $m$
\begin{equation}\label{eq:modzero}
m\equiv0\bmod {p^k} \Leftrightarrow \forall i <k\left[{m \choose p^i}\equiv0\bmod p\right].
\end{equation} 
Define the $i$th elementary symmetric function of the $n$ variables $y_1,\ldots,y_n$, $i\leq n$, as
$$\sum_{1\leq \ell_1 < \cdots < \ell_i \leq n}\prod_{j=1}^i y_{\ell_j}.$$
Note that if each $y_i \in\01$, and exactly $|y|$ of them are 1, then the value of the above
expression is ${|y| \choose i}$. Now write $g$ as a sum of monomials of coefficient 1, i.e., replace
for example $3x_1x_2$ by $x_1x_2 + x_1x_2 + x_1x_2$. Let ${g(x) \choose i}$ be the $i$-th elementary 
symmetric function of the monomials in $g$. Define $h(x)$ as 
$$h(x) = \sum_{i=0}^{k-1}{g(x) \choose p^i} \prod_{j=0}^{i-1}\left(1-{g(x) \choose p^j}^{p-1}\right).$$
The degree of $h(x)$ is $d(2p^{k-1}-1)$. If $g(x)\equiv0 \bmod p^k$, then by Equation \ref{eq:modzero},
${g(x) \choose p^i} \equiv 0 \bmod p$ for all $0\leq i < k$, hence $h(x)\equiv 0 \bmod p$. On the
other hand, if $g(x) \not\equiv 0 \bmod p^k$, then using Equation \ref{eq:modzero}, let $r$ be the least 
value such that ${g(x) \choose p^r} \not\equiv 0 \bmod p$. Note that the $r$th term in $h(x)$ is 
nonzero modulo $p$, but all the others are zero modulo $p$, since all terms after the $r$th contain the 
factor $\left(1-{g(x) \choose p^r}^{p-1}\right) = 0$, and hence $h(x)\not\equiv 0 \bmod p$.
\end{Proof}

\section{Linear Lower Bound for Prime Power Moduli}
In this section we restrict ourselves to the field $\mathbb{Z}_{p^k}$, where $p$
is a prime. For a binary string $x$, let $|x|$ denote its Hamming weight (number of 1's).
We consider any Boolean function $g:\01^n\to\01$, that has $g(x)=1$ if $|x|=n/4$ and $g(x)=0$ 
if $|x|=3n/4$, and prove that it has $\deg(g,p^k)=\Omega(n)$.

The rough idea behind our proof is the following. Using Lemmas \ref{lem:pol01} and
\ref{lem:ppk} we restrict ourselves to polynomials over $\mathbb{Z}_p$ that are
always 0/1 valued on inputs from the domain $\01^n$. This only increases the degree
by a multiplicative constant. Now assume there exists a low degree polynomial $q$ that
represents $g$. We then use the property that for all $x\in\01^n$, $q(x)\neq0$ if $|x|=n/4$
and $q(x)=0$ if $|x|=3n/4$ to set up a system of linear equations over the coefficients
of the monomials in $q$. Using Theorem  \ref{thm:lucas} we then show that this system is 
unsolvable, and conclude that no such low degree polynomial $q$ exists.
\begin{Theorem}\label{thm:lower}Let $p$ be a prime, $n=4p^r$, and $g:\01^n\to\01$ be such that 
$g(x)=1$ if $|x|=n/4$, and $g(x)=0$ if $|x|=3n/4$. Then 
$$\deg(g,p^k)\geq \frac{n}{4(2p^{k-1}-1)(p-1)}.$$\end{Theorem}
\begin{Proof}We will prove the lemma for primes $p>3$. The case where 
$p \in \{2,3\}$ has an identical proof, and we leave this to the reader.

Consider any degree $d < \frac{n}{4(2p^{k-1}-1)(p-1)}$ multilinear polynomial $p$ over $\mathbb{Z}_{p^k}$
that represents $g$. Using Lemmas
\ref{lem:pol01} and \ref{lem:ppk}, transform $p$ into a polynomial $q$ that represents $g$ over $\mathbb{Z}_p$, and 
that has $q(x)\in\01$ for all $x\in\01^n$. This will only increase the degree of $q$ by a multiplicative factor 
$(p-1)(2p^{k-1}-1)$. We now prove a lower bound of $n/4$ on the degree of $q$.
Write $q$ as $$q(x_1,\ldots,x_n) = \sum_{S \subseteq [n],|S|<n/4} c_S \cdot \mon(S),$$
where $\mon(S)=\prod_{i \in S}x_i$, $|S|$ denotes the size of $S$, and each $c_S \in \mathbb{Z}_p$. On input
$x\in\{0,1\}^n$, with $x_1 = \ldots = x_{3n/4}=1$ and $x_{3n/4+1}=\ldots=x_n=0$, 
we have that 
\begin{equation}\label{eq:constraint1}\sum_{S\subseteq [3n/4],|S|<n/4} c_S \equiv 0\bmod p.\end{equation}
However, every input $x\in\{0,1\}^n$ with exactly $n/4$ out of the first 
$3n/4$ variables set to 1 gives a constraint
$$\sum_{S \subseteq T}c_S \equiv 1\bmod p,$$
where $T\subseteq [3n/4]$, is the set of $n/4$ indices of variables that are set to 1 in the 
input $x$. Note that the total number of such constraints modulo $p$ is 
$${3n/4 \choose n/4}\bmod p ={3p^r \choose p^r}\bmod p =3,$$
by Theorem \ref{thm:lucas}. Also, note that every monomial $\mon(S)$ with $S \subseteq
[3n/4]$, of degree $0 < \ell < n/4 = p^r$ occurs in exactly 
$${3n/4 -\ell \choose n/4-\ell}\bmod p={3p^r-\ell \choose p^r-\ell}\bmod p = 1,$$
constraints modulo $p$, which follows again from Theorem \ref{thm:lucas}. To see this, note that
the first $r$ digits in the $p$-ary representation of $3p^r-\ell$ and $p^r-\ell$ for $0<\ell<p^r$
are all equal, but the $(r+1)$st digit of $(p^r-\ell)_p$ is 0, and that of $(3p^r-\ell)_p$ is
2.

Hence summing all these constraints gives
$$2c_\emptyset + \sum_{S\subseteq[3n/4],|S|<n/4} c_S \equiv 3\bmod p,$$
where the term $2c_\emptyset$ is due to the fact that the free term $c_\emptyset$ of $q$ occurs in 
3 constraints modulo $p$. Since $c_\emptyset \in \{0,1\}$ (because $q(0^n)\in\01$), we thus have a 
contradiction with Equation \ref{eq:constraint1}. Hence $q$ must have degree $\geq n/4$.
\end{Proof}
As a consequence of Theorem \ref{thm:lower} we have a relativized separation of $\EQP$ from $\MODpkP$.
\begin{Corollary}There exists an oracle $A$, such that $$\EQP^A \nsubseteq \MODpkP^A.$$ 
\end{Corollary}
\begin{Proof}
For fixed $r$, define $r^* = \lceil \log_2 4p^r\rceil$, and for each $A\subseteq\01^*$,
define $A^{r^*}$ to be the restriction of $A$ to the lexicographically first $4p^r$ strings
of length $r^*$. Consider oracles $A$ with the property that $|A^{r^*}| \in \{p^r,3p^r\}$ for all 
$r$. For such $A$, define $$L_A = \{0^r \mid |A^{r^*}|=p^r\}.$$

Grover's algorithm \cite{grover:search} has the property that if either a $1/4$ or a $3/4$ 
fraction of the total search space is a solution, then we can find out which of the two
is the case with certainty using just one query. This observation was first made by Boyer
\etal \cite{bbht:tight}, and later generalized by Brassard \etal \cite{bhmt:countingj}. Using
this observation, it is not hard to see that for all appropriate $A$, $L_A \in \EQP^A$.

We now show the existence of an $A$ such that $L_A \notin \MODpkP^A$. The construction of $A$ 
will be in stages. Let $M_1,M_2,\ldots$ be an enumeration of $\MODpkP$ oracle machines.  In 
stage $i$, run $M_i$ on input $0^{r_i}$, where $r_i$ is chosen large enough as not to interfere 
with any previous stages. Note that we may assume that $M_i$ only makes queries to the lexicographically first 
$4p^{r_i}$ strings of length $r_i^*$. Call these strings $y_1,y_2,\ldots,y_{4p^{r_i}}$. Take an arbitrary computation
path of $M_i$ on input $0^{r_i}$, and let $y_{i_1},y_{i_2},\ldots,y_{i_\ell}$ be the queries 
made along this path. Note that $\ell$ is upper bounded by a polynomial in $r_i$. Now for each
possible appropriate setting of $A$ on length $r_i^*$, see if this path accepts. If it does, create a
monomial which is the product of all the variables $y_i$ (if $y_i \in A$) or $(1-y_i)$ (if
$y_i \notin A$), for $1\leq i \leq \ell$. Repeat this procedure for all other computation paths.
The sum of all monomials thus obtained is a polynomial $q:\mathbb{Z}_{p^k}^{4p^{r_i}}\to
\mathbb{Z}_{p^k}$, that for $x\in\01^{4p^{r_i}}$ has $q(x)\neq0$ if $|x|=p^{r_i}$, and
$q(x)=0$ if $|x|=3p^{r_i}$. Furthermore, the degree of $q$ is bounded by a polynomial in $r_i$. But
Theorem \ref{thm:lower} states that such a polynomial does not exist. Hence there must
exist a setting of $A$ on length $r^*_i$ such that $M_i$ is incorrect on input $0^{r_i}$.
Set $A$ in this way on length $r^*_i$, this ensures that $M_i$ can not decide $L_A$. Continue with stage $i+1$.
\end{Proof}
In the other direction, an oracle separation of $\MODpkP$ from $\EQP$ is easy to achieve. 
For instance, to separate $\MOD2P$ ($=\PARP$) from $\EQP$, we can use the following construction.
Let $B\subseteq\01^*$, and define $$L_B = \{0^r \mid \textrm{the parity of the number of
strings in $B$ of length $r$ is odd}\}.$$ Clearly, $L_B\in\MOD2P^B$.
However, using the fact that the degree over the reals of the representing polynomial for the 
parity function on $n$ variables is $n$ (Beals \etal \cite{bbcmw:polynomials}), we can show
that there exists a $B$ such that $L_B \notin \EQP^B$.

\section{Sublinear Upper Bound for Squarefree Composite Moduli}
We now focus our attention on representing $g$ over the ring $\mathbb{Z}_m$ of
integers modulo $m$, where $m$ is a squarefree composite. We prove the following theorem.
\begin{Theorem}\label{thm:uppersf}Let $m$ be a squarefree composite 
with largest prime factor $p$, $4\mid n$, and $g:\01^n\to\01$ be such that $g(x)=1$ if $|x|=n/4$, and $g(x)=0$ 
if $|x|=3n/4$. Then $\deg(g,m)=\mathcal{O}(n^{1/\log_p m})$.\end{Theorem}
We will prove this result using two separate lemmas. Note that $g$ is only well-defined on
lengths $n$ such that $4\mid n$. We split these possible lengths in 2 different categories,
namely those such that $n/4 \not\equiv 0\bmod m$, and those such that $n/4 \equiv 0 \bmod m$.
For each of these two lengths, we separately prove an upper bound. 
A key insight that we shall need is provided by the following lemma.
\begin{Lemma}\label{lem:digits}If $m=p_1p_2\cdots p_r$, then for all $1 \leq i \leq r$, $${am^b \choose p_i^\ell} \bmod m \neq 
{3am^b \choose p_i^\ell}\bmod m$$ if and only if the $(\ell+1)$st digit of the $p_i$-ary representations of 
$am^b$ and $3am^b$ differ.\end{Lemma}
\begin{Proof}Let $1 \leq j \leq r$ with $j\neq i$. We have that $p_j \nmid p_i^\ell$, but $p_j \mid a\cdot m^b$. Hence
the first digit of $(am^b)_{p_j}$ is 0 and the first digit of $(p_i^\ell)_{p_j}$ is nonzero. By Theorem 
\ref{thm:lucas}, it follows that  $${am^b \choose p_i^\ell} \bmod p_j = 0,$$
for $1\leq j\leq r$, and $j\neq i$. Note that again by Theorem \ref{thm:lucas}, the value of ${am^b \choose p_i^\ell}
\bmod p_i$ is determined by the $(\ell+1)$st digit of $(am^b)_{p_i}$, since the only nonzero
digit of $(p_i^\ell)_{p_i}$ is the $(\ell+1)$st and has value 1. Likewise, the value of ${3am^b 
\choose p_i^\ell}$ is determined by the $(\ell+1)$st digit of $(3am^b)_{p_i}$.
Now apply the Chinese Remainder Theorem.\end{Proof}
The use of Lemma \ref{lem:digits} stems from the following fact. Assume we have a polynomial
$p:\mathbb{Z}_m^n \to \mathbb{Z}_m$, where all the monomials of degree $d$ have coefficient 1,
and all other monomials have coefficient 0. Then for $x \in\01^n$, $p(x)$ has the value ${|x| \choose d}\bmod m$.
Specifically, if ${n/4 \choose d} \bmod m \neq {3n/4 \choose d} \bmod m$, then $p(x)-{3n/4 \choose d}$
is a representing polynomial for $g$ of degree $d$. 

We shall first prove that if $n/4=am^b$ for some $b>0$, and $0<a<m$, then there exists a degree at 
most $p^{b+1}=\mathcal{O}(n^{1/\log_pm})$ representing polynomial for $g$ over $\mathbb{Z}_m$, where 
$p$ is the least prime factor of $m$ not occurring in $a$. 
\begin{Lemma}\label{lem:upper1}Let $m=p_1p_2\cdots p_r$ be a squarefree composite with 
$p_i < p_{i+1}$, $0<a<m$, $b>0$, and $p$ the least prime factor of $m$ not occurring in $a$. Then the following hold.
\begin{enumerate}
\item if $p > 2$, then ${am^b \choose p^b} \bmod m \neq {3am^b \choose p^b} \bmod m$
\item if $p = 2$, then ${am^b \choose p^{b+1}} \bmod m \neq {3am^b \choose p^{b+1}} \bmod m$
\end{enumerate}
\end{Lemma}
\begin{Proof}To prove item 1, we will show that for all $b$, the $(b+1)$st digit of 
$(am^b)_{p}$ and $(3am^b)_{p}$ are different. The result then follows by Lemma 
\ref{lem:digits}. We distinguish between the case where $p=3$, and $p>3$.

If $p = 3$, then since for all $i \leq b$, $3^i \mid am^b$, 
but $3^{b+1} \nmid am^b$, the $(b+1)$st digit of $(am^b)_3$ is the first nonzero digit. 
However, since $3^i \mid 3am^b$ for $i \leq b+1$, the $(b+1)$st digit of $(3am^b)_3$ is 0.

If $p > 3$, then
for all $0 < i \leq b$, we have that $p^i \mid am^b$, and $p^i \mid 3am^b$, but $p^{b+1}\nmid am^b$
and $p^{b+1} \nmid 3am^b$.
Hence, we have that the $(b+1)$st digit of both $(am^b)_{p}$ and $(3am^b)_{p}$ 
is the first nonzero digit. We now claim that $(am^b \bmod p^{b+1})/p^b
\neq (3am^b \bmod p^{b+1})/p^b$, i.e., $(am^b)_{p}$ and $(3am^b)_{p}$ differ
in their $(b+1)$st digit. To prove this, note that $am^b = c\cdot p^{b+1} + 
r\cdot p^b$, for some integer $c$ and $0 < r < p$. Hence, 
$3am^b = c'\cdot p^{b+1}+ 3r\cdot p^b$. But $3r \bmod p \neq r$, for 
all $0 < r < p$, if $p>3$ and $p$ is a prime. 

To prove item 2, we show that the $(b+2)$nd digit of $(am^b)_2$ and $(3am^b)_2$
differ, for all $b$. Note that for both $(am^b)_2$ and $(3am^b)_2$, the first nonzero bit 
is the $(b+1)$st, so both $(am^b)_2$ and $(3am^b)_2$ are of the form $\ldots10^b$. Now
if $(am^b)_2$ has a 0 as its $(b+2)$nd bit, i.e. $(am^b)_2$ is of the form $\ldots010^b$, then
$(3am^b)_2$ is of the form $\ldots110^b$. On the other hand, if $(am^b)_2$ has a 1 as its
$(b+2)$nd bit, i.e. $(am^b)_2$ is of the form $\ldots110^b$, then $(3am^b)_2$ is of the form 
$\ldots010^b$. Hence the $(b+2)$nd bit of $(am^b)_2$ and $(3am^b)_2$ are different.\end{Proof}
If $n$ is such that $n/4\not\equiv0 \bmod m$, then we can prove that constant degree suffices.
\begin{Lemma}\label{lem:upper2}Let $m=p_1p_2\cdots p_r$ be a squarefree composite with 
$p_i < p_{i+1}$. Then the following hold.
\begin{enumerate}
\item if $c\neq m/2$, then ${am+c \choose 1} \bmod m \neq {3am+3c \choose 1} \bmod m$, for $0<c<m$
\item if $c=m/2$, then ${am+m/2 \choose 2} \bmod m \neq {3am+3m/2 \choose 2} \bmod m$
\end{enumerate}
\end{Lemma}
\begin{Proof}We first prove item 1. If $c \neq m/2$, then $c\not\equiv 3c \bmod m$. Hence, 
$(am+c) \bmod m \neq (3am + 3c) \bmod m$.

We now prove item 2. Since $m$ is an even squarefree number, $m/2$ is odd. Hence the first bit of
$(am+m/2)_2$ is 1. We thus have that $(am+m/2)_2$ has the form $\ldots11$ or $\ldots01$. In the
first case, $(3am+3m/2)_2$ then has the form $\ldots01$, in the second case, $(3am+3m/2)_2$ has
the form $\ldots11$. In other words, $(am+m/2)_2$ and $(3am+3m/2)_2$ differ in their 2nd bit.
Using Theorem \ref{thm:lucas} and the Chinese Remainder Theorem, the result follows.\end{Proof}
\vbox{Together, Lemmas \ref{lem:upper1} and \ref{lem:upper2} imply Theorem \ref{thm:uppersf}:
\begin{Proof}{\bf (of Theorem \ref{thm:uppersf})} Let $m$ be a squarefree composite with largest
prime factor $p$, $4\mid n$, and $g:\01^n\to\01$ be such that $g(x)=1$ if $|x|=n/4$, and $g(x)=0$
if $|x|=3n/4$. We will exhibit a representing polynomial of degree $\mathcal{O}(n^{1/\log_pm})$ 
for $g$ on each length $n$. We distinguish two different cases for $n$:
\begin{enumerate}
\item $n/4 \equiv 0 \bmod m$, i.e., $n/4=am^b$ with $0<a<m$, and $b>0$. In this case, Lemma 
\ref{lem:upper1} tells us that if $p$ is the least prime factor of $m$ not in $a$, then
either ${am^b \choose p^b} \bmod m \neq {3am^b \choose p^b} \bmod m$ (if $p>2$), or 
${am^b \choose p^{b+1}} \bmod m \neq {3am^b \choose p^{b+1}} \bmod m$ (if $p=2$). In the former
case, the polynomial $${3am^b \choose p^b} - \sum_{S\subseteq[n],|S|=p^b} \mon(S),$$
where $\mon(S)=\prod_{i\in S}x_i$, represents $g$. This polynomial has degree $p^b$. In the latter case $${3am^b \choose p^{b+1}} - 
\sum_{S\subseteq[n],|S|=p^{b+1}} \mon(S)$$ is a representing polynomial of degree $p^{b+1}$. 
\item $n/4 \not\equiv 0 \bmod m$, i.e., $n/4=am+c$. In this case Lemma \ref{lem:upper2}
tells us that either $am+c \bmod m \neq 3am+3c \bmod m$ (if $c\neq m/2$), or ${am+c \choose 2}
\bmod m \neq {3am+3c \choose 2} \bmod m$ (if $c=m/2$). In the former case, the polynomial
$$3am+3c - \sum_{i=1}^n x_i$$ is a representing polynomial for $g$ of degree 1. In the latter
case, $${3am+3c \choose 2} - \sum_{S\subseteq[n],|S|=2}\mon(S)$$ is a degree 2 representing
polynomial for $g$.
\end{enumerate}
\end{Proof}}

\section{Discussion and Open Problems}
We studied the degree of a polynomial $q:\mathbb{Z}_m^n\to\mathbb{Z}_m$, that for all $x\in\01^n$
has $q(x)\neq0$ if $|x|=n/4$, and $q(x)=0$ if $|x|=3n/4$. We have proven a linear lower bound when 
$m$ is a prime power, and an upper bound of $\mathcal{O}(n^{1/\log_p m})$, if $m$ is a squarefree 
composite with largest prime factor $p$. The former result implies a relativized separation of
$\EQP$ from $\MODpkP$.

A number of open questions are left by this research. First of all, can we prove that the upper
bound of $\mathcal{O}(n^{1/\log_p m})$ is tight? And second, what can we say about general
composite $m$, instead of \emph{squarefree} composite $m$? Establishing a good lower bound in the
latter case would show a relativized separation of $\EQP$ from $\MODmP$ for all $m$. 

Another interesting direction is to investigate whether one can exhibit an oracle relative to 
which $\EQP$ is not contained in $\SI2$ or higher levels
of $\PH$. This will require, however, a different and presumably more complex oracle construction 
than the one we have used here, since the language that separates $\EQP^A$ from $\PNP^A$ and 
$\MODpkP^A$ is in $\BPP^A$, and hence in $\SI2^A$.

\section*{Acknowledgments}
The first author would like to thank Harry Buhrman, Frederic Green, Leen Torenvliet, and 
Ronald de Wolf for some interesting and useful discussions on the subject. 
The second author would like to thank Madhu Sudan for bringing this
problem to his attention.


\begin{thebibliography}{BHMT00}

\bibitem[BB94]{bb:oracle}
A.~Berthiaume and G.~Brassard.
\newblock Oracle quantum computing.
\newblock {\em Journal of Modern Optics}, 41(12):2521--2535, 1994.

\bibitem[BBC{\etalchar{+}}98]{bbcmw:polynomials}
R.~Beals, H.~Buhrman, R.~Cleve, M.~Mosca, and R.~{de} Wolf.
\newblock Quantum lower bounds by polynomials.
\newblock In {\em Proceedings of 39th IEEE FOCS}, pages 352--361, 1998.
\newblock quant-ph/9802049.

\bibitem[BBHT98]{bbht:tight}
M.~Boyer, G.~Brassard, P.~H{\o}yer, and A.~Tapp.
\newblock Tight bounds on quantum searching.
\newblock {\em Fortsch. Phys.}, 46:493--506, 1998.

\bibitem[BBR92]{bbr:modulo}
D.~Mix Barrington, R.~Beigel, and S.~Rudich.
\newblock Representing {Boolean} functions as polynomials modulo composite
  numbers.
\newblock In {\em Proceedings of the 24th ACM Symposium on Theory of
  Computing}, pages 455--461, 1992.

\bibitem[Bei93]{beigel:poly}
R.~Beigel.
\newblock The polynomial method in circuit complexity.
\newblock In {\em Proceedings of the 8th {IEEE} Structure in Complexity Theory
  Conference}, pages 82--95, 1993.

\bibitem[BHMT00]{bhmt:countingj}
G.~Brassard, P.~H{\o}yer, M.~Mosca, and A.~Tapp.
\newblock Quantum amplitude amplification and estimation.
\newblock quant-ph/0005055. This is the upcoming journal version
  of~\cite{bht:counting,mosca:eigen}, 15 May 2000.

\bibitem[BHT98]{bht:counting}
G.~Brassard, P.~H{\o}yer, and A.~Tapp.
\newblock Quantum counting.
\newblock In {\em Proceedings of 25th ICALP}, volume 1443 of {\em Lecture Notes
  in Computer Science}, pages 820--831. Springer, 1998.
\newblock quant-ph/9805082.

\bibitem[BV97]{bernstein&vazirani:qcomplexity}
E.~Bernstein and U.~Vazirani.
\newblock Quantum complexity theory.
\newblock {\em SIAM Journal on Computing}, 26(5):1411--1473, 1997.
\newblock Earlier version in STOC'93.

\bibitem[GP01]{green&pruim:eqp_pnp}
F.~Green and R.~Pruim.
\newblock Relativized separation of {EQP} from {P(NP)}.
\newblock {\em Information Processing Letters}, 80(5):257--260, 2001.

\bibitem[Gro96]{grover:search}
L.~K. Grover.
\newblock A fast quantum mechanical algorithm for database search.
\newblock In {\em Proceedings of 28th ACM STOC}, pages 212--219, 1996.
\newblock quant-ph/9605043.

\bibitem[Luc78]{lucas:congruences}
E.~Lucas.
\newblock Sur les congruences des nombres eul{\'e}riens et les coefficients
  diff{\'e}rentiels des fonctions trigonom{\'e}triques, suivant un module
  premier.
\newblock {\em Bull. Soc. Math. France}, 6:49--54, 1878.

\bibitem[Mos98]{mosca:eigen}
M.~Mosca.
\newblock Quantum searching, counting and amplitude amplification by
  eigenvector analysis.
\newblock In {\em MFCS'98 workshop on Randomized Algorithms}, pages 90--100,
  1998.

\bibitem[NC00]{nielsen&chuang:qc}
M.~A. Nielsen and I.~L. Chuang.
\newblock {\em Quantum Computation and Quantum Information}.
\newblock Cambridge University Press, 2000.

\bibitem[Pap94]{papadimitriou:cc}
C.~H. Papadimitriou.
\newblock {\em Computational Complexity}.
\newblock Addison-Wesley, 1994.

\bibitem[TB95]{tardos:lower}
G.~Tardos and D.~Mix Barrington.
\newblock A lower bound on the mod 6 degree of the {OR} function.
\newblock In {\em Israel Symposium on Theory of Computing Systems}, pages
  52--56, 1995.

\end{thebibliography}

\newcommand{\etalchar}[1]{$^{#1}$}

\end{document}